\documentclass[twocolumn,showpacs,preprintnumbers,amsmath,amssymb,prl,epsf]{revtex4}
\usepackage{graphicx}

\begin{document}
\title{Multiphoton ghost imaging with classical light}
\author{I.~N.~Agafonov, M.~V.~Chekhova, A.~N.~Penin}
\affiliation{Department of Physics, M.V.Lomonosov Moscow State
University,\\  Leninskie Gory, 119992 Moscow, Russia} \vskip 24pt

\begin{abstract}
\begin{center}\parbox{14.5cm}
{One of the possible types of n-th order ghost imaging is
experimentally performed using multi-photon (higher-order) intensity
correlations of pseudothermal light. It is shown that although
increasing the order of intensity correlations leads to the growth
of ghost imaging visibility, it at the same time reduces the
signal-to-noise ratio. Therefore, ghost imaging with thermal light
is optimal in the second order in the intensity.}
\end{center}
\end{abstract}
\pacs{42.50.Dv, 03.67.Hk, 42.62.Eh}
\maketitle \narrowtext
\vspace{-10mm}

\section{Introduction}

The technique of ghost imaging (GI), first demonstrated in
1995~\cite{Shih+Pittman}, still attracts much interest in connection
with remote sensing~\cite{Meyers}, lenseless imaging~\cite{Han}, and
fundamental problems, such as the boundary between quantum and
classical physics (see, for instance,~\cite{Erkmen1}). The idea of
ghost imaging is based on the correlations existing between the
intensities of two beams, which will be further called the signal
one and the reference one. The object to be imaged (a mask) is
placed into the signal beam, and the photons transmitted through it
are registered by a ``bucket'' detector, which collects all signal
radiation and hence does not resolve the shape of the mask. The
detector in the reference beam, on the contrary, is spatially
resolving, but there is no mask in the reference beam. The
coincidences, or photocurrent correlations, of signal and reference
detectors are registered as a function of the reference detector
position. When the reference detector registers photons whose
correlated signal counterparts pass through the mask, there is an
increase in the coincidence counting rate (photocurrent correlation
function). Thus, by scanning the reference detector one restores the
image of the mask in the signal beam.

Although ghost imaging was at first supposed to require two-photon
light generated via spontaneous parametric down-conversion (SPDC),
it was later demonstrated with classical light~\cite{Boyd}, in
particular, pseudo-thermal light obtained by passing a coherent beam
through a rotating ground-glass disk~\cite{Lugiato,Shih,Zhang} and
true thermal light~\cite{LA2009}. In a large number of works, both
near-field and far-field ghost imaging with pseudo-thermal light was
performed, as well as ghost imaging of a purely phase
object~\cite{Lugiatophase} or a reflecting object~\cite{Meyers}. It
seemed that the only advantage of using nonclassical light for ghost
imaging was a higher visibility than in the case of thermal light.
However, in several works it was proposed to increase the visibility
of ghost imaging with classical light by increasing the value of
normalized correlation function. In Ref.~\cite{Basano}, it was
performed by introducing a threshold for the bucket detector output
and hence choosing only peaks of the intensity fluctuations. In
Ref.~\cite{ghostJMO}, it was suggested to use the fact that the
normalized intensity correlation function (CF) of the $n$-th order
for thermal light grows as $n!$ and thus to increase the visibility
of ghost imaging with thermal light by passing to higher-order
intensity correlations. This idea was realized recently~\cite{LA}
and the growth of ghost-imaging visibility with the order of the
registered intensity moment was indeed demonstrated in experiment.

However, as it was pointed out in
Refs.~\cite{ghostJMO,Erkmen2,Boyd}, it is not the visibility of a
ghost image that matters but rather its signal-to-noise ratio (SNR).
The noise of a ghost image is related to the variance of the
intensity moment under measurement and contains both the shot-noise
component and the excess noise component, as well as intermediate
terms~\cite{Erkmen2}. Theoretical quantum calculation of the noise
for ghost imaging with thermal and nonclassical light has been
performed in Ref.~\cite{Erkmen2}; in Ref.~\cite{Basano2} the noise
for ghost imaging with pseudo-thermal light was also measured
experimentally. In Ref.~\cite{Boyd} the so called
``contrast-to-noise ratio'' has been theoretically studied for
high-order thermal ghost imaging using a classical approach.
However, the present work is the first consideration and
experimental measurement of SNR for multiphoton ghost imaging with
pseudo-thermal light, i.e., ghost imaging based on higher-order
intensity moments.

\section{Theory}
There are many ways to generalize ghost imaging in terms of using
higher order intensity correlations. First for the sake of
simplicity we consider an (n-1)-port Hanbury-Brown and Twiss (HBT)
setup depicted in Fig. 1. As in a standard lensless ghost imaging
setup, radiation from a thermal source (usually a rotating
ground-glass disk illuminated by a laser beam) is split on a
beamsplitter (${\textrm{BS}}_1$). One of the two output beams (the
reference one) is further split on n-2 beamsplitters and detected in
the far-field zone by n-1 spatially resolving detectors at locations
${\mathbf{x}}_1,\ldots,{\mathbf{x}}_{n-1}$. The other (signal) beam
is sent to a bucket detector (B) with a mask (M) placed in front of
it. Using all n detectors, one measures the intensity correlation
function of order $n$
\begin{equation}
\tilde{G}^{(n)}(\mathbf{x}_1,\ldots,\mathbf{x}_{n-1})\equiv\langle
\prod_{i=1}^{n-1} I(\mathbf{x}_i) \int_{A_s} I(\mathbf{y})
T(\mathbf{y}) \hbox{d}\mathbf{y}\rangle, \label{G_n}
\end{equation}
where $I(\mathbf{x}_i)$ and $I(\mathbf{y})$ are instant intensities
at corresponding locations measured by (n-1) reference and one
signal detectors, respectively. Angle brackets correspond to
time-averaging and transmission of the mask is denoted by the
function $T(\mathbf{y})$ taking only values $0$ or $1$. Because the
bucket detector collects all signal radiation, integration is
implied over its area $A_s$. Since the bucket detector is supposed
to be larger than the mask, integration can be done over the mask
area.

\begin{figure}
\includegraphics[width=0.38
\textwidth]{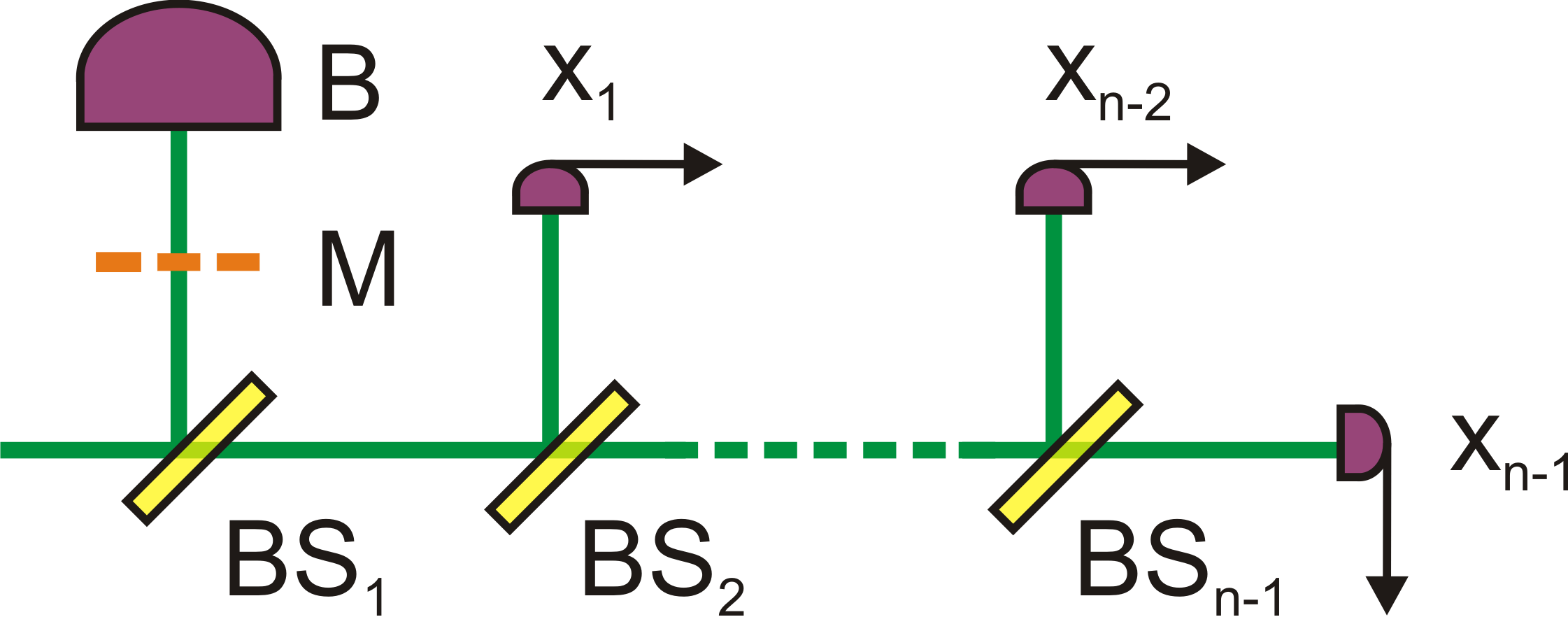} \caption{Schematic of an n-port HBT setup for
measuring an n-point n-th order CF.}
\end{figure}

Since it is rather complicated to analyze properties of an n-point
correlation function (or (n-1)-point ghost image) we focus on a more
simple 2-point case where positions of all (n-1) reference detectors
coincide: $x_1=\ldots=x_{n-1}=x$. Eq. (\ref{G_n}) then turns into
\begin{equation}
G^{(n)}(\mathbf{x})\equiv\langle I^{n-1}(\mathbf{x}) \int_{A_s}
I(\mathbf{y}) T(\mathbf{y}) \hbox{d}\mathbf{y}\rangle. \label{G}
\end{equation}

To simplify the calculation, we will consider one-dimensional case
and use a very convenient approach of Ref.~\cite{Basano2} where a
discrete set of modes was introduced. It means that from continuous
intensity distributions $I(\mathbf{x})$ we pass to the values of
intensities at discrete points $\mathbf{x}_{k}$, and  integration in
Eq.~(\ref{G}) is replaced by summation over the modes $k'$
transmitted by the mask, $1\le k'\le M$:
\begin{equation}
G_k^{(n)}\equiv\langle I_k^{n-1}\sum_{k'=1}^M I_{k'}\rangle.
\label{Gsum}
\end{equation}

Calculation of the intensity correlators from Eq.~(\ref{Gsum}) can
be easily performed taking into account that for thermal light
\begin{equation}
\langle I_k^{l}I_{k'}^{m}\rangle=\{(l+m)!\,\delta_{kk'}+l!\,m!\,(1-\delta_{kk'})\}I^{l+m}.
\label{corr}
\end{equation}
Here, $I$ is the time-averaged intensity, which is assumed to be
uniform over the whole area of the mask, and $\delta_{k,k'}$ is the
Kronecker symbol. Hence, for the positions of the reference detector
that have counterparts in the signal beam transmitted through the
mask, $1\le k\le M$, the correlation function \ref{Gsum} will take
its maximum value,
\begin{equation}
G^{(n)}_{\mathrm{max}}=(n-1)!\,(M+n-1)\,I^n.
\label{corrmax}
\end{equation}

Otherwise, if $k$ is outside the interval $[1,M]$, the correlation
function will take the 'background' value,
\begin{equation}
G^{(n)}_{\mathrm{back}}=(n-1)!\,M\,I^n.
\label{corrmin}
\end{equation}

The visibility of $n$th-order ghost imaging can be defined as
\begin{equation}
V^{(n)}\equiv\frac{G^{(n)}_{\mathrm{max}}-G^{(n)}_{\mathrm{back}}}{G^{(n)}_{\mathrm{max}}+G^{(n)}_{\mathrm{back}}}.
\label{vis}
\end{equation}
From Eqs.(\ref{corrmax},\ref{corrmin}), it becomes
\begin{equation}
V^{(n)}=\left(\frac{2M}{n-1}+1\right)^{-1},
\label{visn}
\end{equation}
i.e., the visibility grows with the order of the correlation
function measured. From Eq.(\ref{visn}), one can see that the
visibility decrease due to a large number $M$ of modes in the image
can be reduced by passing to a high order $n$. In the single-mode
case, $M=1$, the values of visibility for different orders are
$V^{(2)}=1/3$, $V^{(3)}=1/2$, $V^{(4)}=3/5$,... Note that the
results for the maximum and background values of the correlation
function, as well as for the visibility, will be the same if
quantum-mechanical approach is used in the calculation. However, in
the calculation of the noise it is important to use the
quantum-mechanical treatment (as, for instance, in
Ref.~\cite{Erkmen2}, since it is the only way to take into account
the shot noise. For this purpose, we will write the expression for
the maximum and background correlation functions in the quantum
form,

\begin{eqnarray}
G^{(n)}_{\mathrm{max}}\equiv\langle (a_1^{\dagger})^{n-1}\sum_{k=1}^M a_k^{\dagger}a_k\,\,a_1^{n-1}\rangle,\nonumber
\\
G^{(n)}_{\mathrm{back}}\equiv\langle (a_0^{\dagger})^{n-1}\sum_{k=1}^M a_k^{\dagger}a_k\,\, a_0^{n-1}\rangle.
\label{Gquant}
\end{eqnarray}
Here, angular brackets denote averaging over the thermal state of
$M+1$ modes whose photon creation and annihilation operators are
$a_k^{\dagger},a_k$, $k=0,1,\dots M$. The mode $k=0$ is the only one
considered outside of the mask.

We will now calculate the signal-to-noise ratio, taking as the
signal the difference $S\equiv
G^{(n)}_{\mathrm{max}}-G^{(n)}_{\mathrm{back}}$ and as the noise,
the standard deviation of this value, which is
\begin{equation}
\Delta
S=\sqrt{\mathrm{Var}(G^{(n)}_{\mathrm{max}})+\mathrm{Var}(G^{(n)}_{\mathrm{back}})-2\mathrm{Cov}(G^{(n)}_{\mathrm{max}},G^{(n)}_{\mathrm{back}})},
\label{noise}
\end{equation}
where $\mathrm{Var}(G^{(n)}_{\mathrm{max}})$ and
$\mathrm{Var}(G^{(n)}_{\mathrm{back}})$ are variances of the maximum
and background correlation functions and
$\mathrm{Cov}(G^{(n)}_{\mathrm{max}},G^{(n)}_{\mathrm{back}})$ is
their covariance. The latter has to be taken into account since
$G^{(n)}_{\mathrm{max}},G^{(n)}_{\mathrm{back}}$ are not
statistically independent.

Calculation of the values entering Eq.(\ref{noise}) is done by
passing to normally-ordered correlators using Wick's theorem
~\cite{Wick}. For instance, $\mathrm{Var}(G^{(n)}_{\mathrm{back}})$
is calculated as

\begin{eqnarray}
\nonumber
\mathrm{Var}(G^{(n)}_{\mathrm{back}})=\\
\nonumber =\sum_{k=1}^M\sum_{k'=1}^M\langle(a_0^{\dagger})^{n-1}
a_k^{\dagger}a_k a_0^{n-1}(a_0^{\dagger})^{n-1}a_{k'}^{\dagger}a_{k'} a_0^{n-1}\rangle\\
-\left[(n-1)!\right]^2M^2I^{2n}, \label{varback}
\end{eqnarray}
which yields
\begin{eqnarray}
\nonumber
\mathrm{Var}(G^{(n)}_{\mathrm{back}})=\left[(n-1)!\right]^2I^{2n}\times\\
\left[\sum_{i=0}^{n-1}
\frac{(2n-i-2)!I^{-i}}{[(n-i-1)!]^2i!}M(M+1+I^{-1})-M^2\right].
\label{varbres}
\end{eqnarray}

Similarly, calculation of $\mathrm{Var}(G^{(n)}_{\mathrm{max}})$ and
$\mathrm{Cov}(G^{(n)}_{\mathrm{max}},G^{(n)}_{\mathrm{back}})$ leads
to the following expression for the signal-to-noise ratio:
\begin{widetext}
\begin{eqnarray}
\nonumber \mathrm{SNR}=(n-1)\left[ \sum_{i=0}^{n-1}
\frac{(2n-i-2)!I^{-i}}{[(n-i-1)!]^2 i!}\left\{
\frac{n^2(2n-i)(2n-i-1)}{(n-i)^2}
+\frac{2(M-1)n(2n-i-1)}{n-i}+2M^2+\frac{2M-1}{I}\right\} + \right.\\
\left.+\frac{n^2}{I^n}+\frac{2(1-M-n^2)}{I}-(M-1)(M+4n-1)-M^2-3n^2\right]
^{-\frac{1}{2}}. \label{SNR}
\end{eqnarray}
\end{widetext}

In the low-intensity limit, this expression becomes
\begin{equation}
\mathrm{SNR|_{I\rightarrow 0}}=\frac{I^{n/2}(n-1)}{2M-1+n^2}.
\label{SNRsmall}
\end{equation}

We see that the signal-to-noise ratio grows polynomially with the
intensity of light, the growth index being $n/2$. By plotting
dependencies (\ref{SNR}) for various orders $n$ for single-mode case
(Fig. 2), we see that although the growth is faster for higher $n$,
the signal-to-noise ratio saturates at large intensities, and the
maximal values of SNR at higher orders are smaller. Indeed, the
high-intensity asymptotic values of SNR are
\begin{eqnarray} \nonumber
\mathrm{SNR|_{I\rightarrow\infty}} = (n-1)[\frac{2(2n-2)!}{[(n-1)!]^2}\times\\
\nonumber \{(2n-1)(n+M-1)+M^2\}-\\
-2(M-1)(M+2n)-3n^2-1]^{-\frac{1}{2}}, \label{SNRlarge}
\end{eqnarray}

i.e., they decrease at large $n$. Analogous calculation for the case
of SPDC gives the following expression for SNR
\begin{equation}
\mathrm{SNR_{SPDC}} = \frac{\sqrt{m (m + 1)}}{\sqrt{1 + 7m + 7m^2 +
2M m (3m + 2) + 2M^2m^2}} \label{SNR_SPDC},
\end{equation}

where $m$ is the mean number of photons per mode. In a
high-intensity limit
\begin{equation}
\mathrm{SNR_{SPDC}|_{m\rightarrow\infty}} =
\frac{1}{\sqrt{7+6M+2M^2}} \label{SNR_SPDC_lim},
\end{equation}
which coincides with the result obtained for second-order thermal
ghost imaging.

\begin{figure}
\includegraphics[height=5cm]{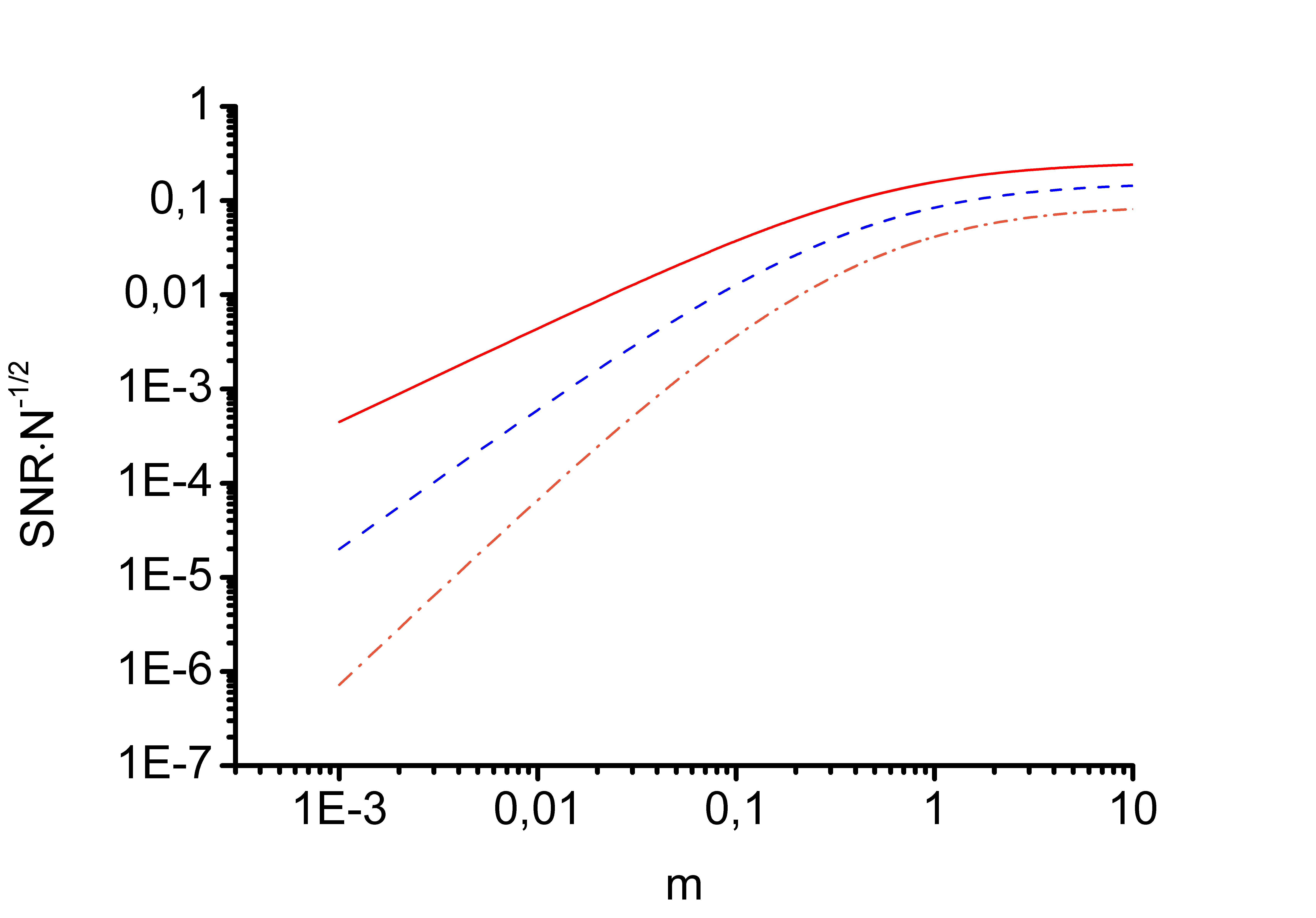}
\caption{Normalized SNR dependence on intensity expressed in photons
per mode for 2-nd (red solid), 3-rd (blue dashed) and 4-th (orange
dot-dashed) order ghost images for single-mode case ($M=1$).}
\end{figure}

Fig. 3 shows the dependence of the normalized SNR on the intensity
expressed in photons per mode for SPDC and second-order thermal
light ghost imaging for single-mode and $M=10$ cases. As one can see
SPDC results in higher SNR values only at low intensities, reaching
a maximum value of about 0.27 at intensity of approximately 0.8
photons per mode for the single-mode case. This maximum is just 6\%
higher that the hight-intensity limit (which is $\frac{1}{\sqrt{15}}
\approx 0.26$), so for single-mode case the difference between the
maximum SNR obtained with biphotons and the one obtained with
thermal light is negligible. For $M=10$ SPDC has a maximum SNR value
of 0.11 at 0.07 photons per mode which is 90\% higher that the
high-intensity limit (which is $\frac{1}{\sqrt{267}} \approx 0.06$
for this case).

\begin{figure}
\includegraphics[width=0.4
\textwidth]{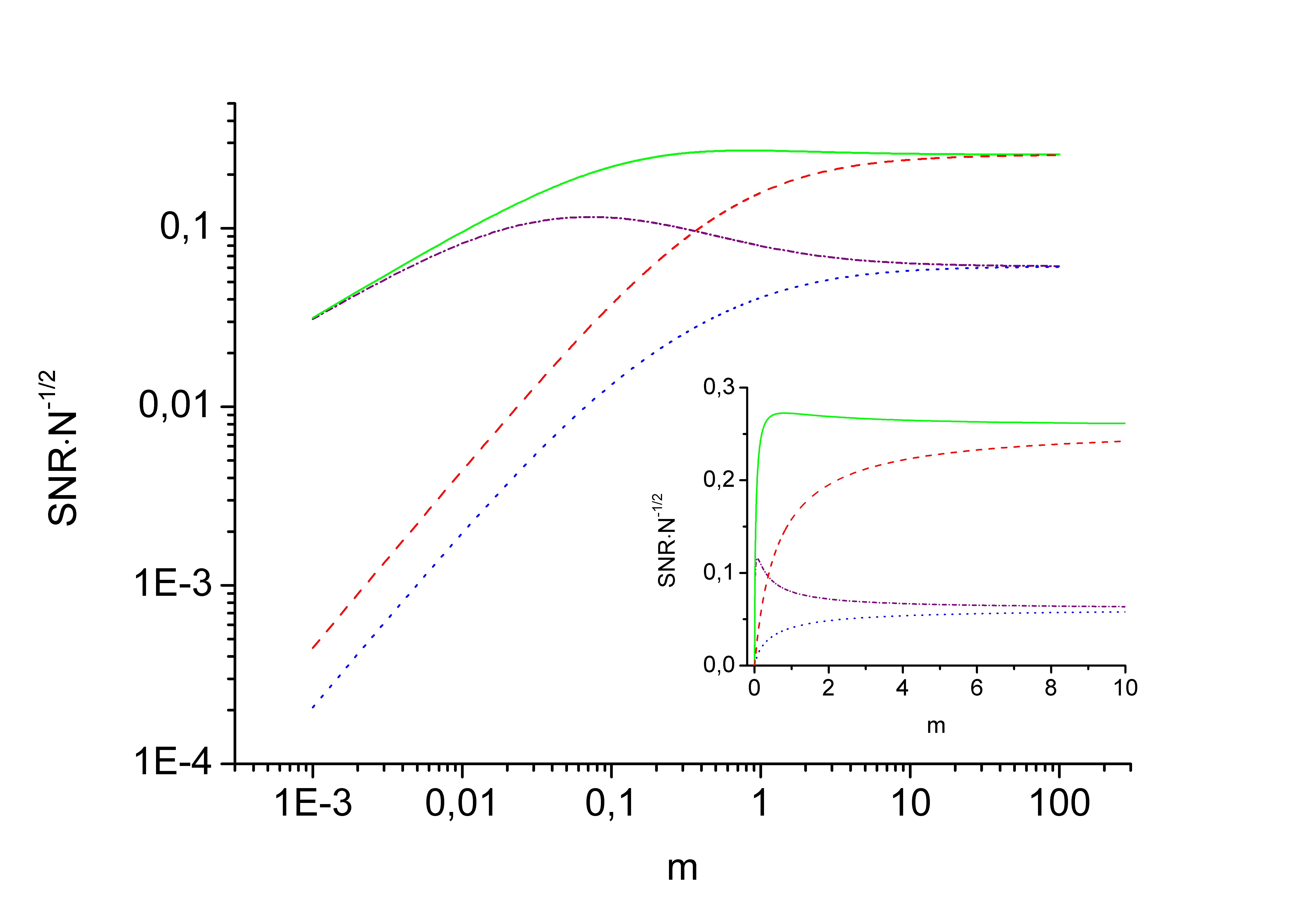} \caption{Normalized SNR dependence on
intensity expressed in photons per mode for single-mode SPDC ghost
imaging (green solid), thermal light GI (red dashed), SPDC ghost
imaging for $M=10$ (purple dash-dotted) and thermal light GI for
$M=10$ (blue dotted). Given in the inset is the central part plotted
in linear scale.}
\end{figure}

Experimentally it is more convenient to measure a 2-point
correlation function using a 2-port HBT setup (Fig. 4) instead of an
n-port setup. Below, we provide a proof of equivalence of the above
mentioned setups for strong thermal light (which is the case of
consideration) in terms of measuring a 2-point n-th order intensity
CF. In quantum description, an n-point n-th order CF is defined by
the following expression
\begin{equation}
G^{(n)}\left( x_1, x_2, \ldots, x_n \right) = \left<
\hat{a}^{\dag}_1 \hat{a}^{\dag}_2 \ldots \hat{a}^{\dag}_n \hat{a}_1
\hat{a}_2 \ldots \hat{a}_n \right>,
\end{equation}
 where $\hat{a}^\dag_i$ and
$\hat{a}_i$ are the photon creation and annihilation operators at
point $x_i$. If n-1 spatial arguments of the CF coincide
$x_1=x_2=\ldots=x_{n-1}$, it becomes
\begin{equation}
G^{(n)}\left( x_1, x_1, \ldots, x_1, x_n \right) = \left< \left(
\hat{a}^{\dag}_1 \right)^{n-1} \hat{a}^{\dag}_n \hat{a}_1^{n-1}
\hat{a}_n \right>.\label{CF_theor}
\end{equation}

The coincidence counting rate measured by the setup shown in Fig. 4
corresponds to a different 2-point CF, having the form
\begin{equation}
\tilde{G}^{(n)}\left( x_1, x_1, \ldots, x_1, x_n \right) = \left<
\hat{a}^{\dag}_1 \hat{a}_1 \hat{a}^{\dag}_1 \hat{a}_1 \ldots
\hat{a}^{\dag}_n \hat{a}_n \right>. \label{CF_meas}
\end{equation}
Using the commutation relations one can represent CF (\ref{CF_meas})
as a series of normally-ordered CFs ~\cite{Wick}
\begin{equation}
\tilde{G}^{(n)}\left( x_1, x_1, \ldots, x_1, x_n \right) =
\sum_{i=2}^n C_i G^{(i)} \left( x_1, \ldots, x_n \right),
\label{CF_eq}
\end{equation}
with $C_n=1$. Each CF $G^{(i)}$ contains 2i-2 operators taken at
point $x_1$ and two operators taken at point $x_n$.

\begin{figure}
\includegraphics[width=0.3
\textwidth]{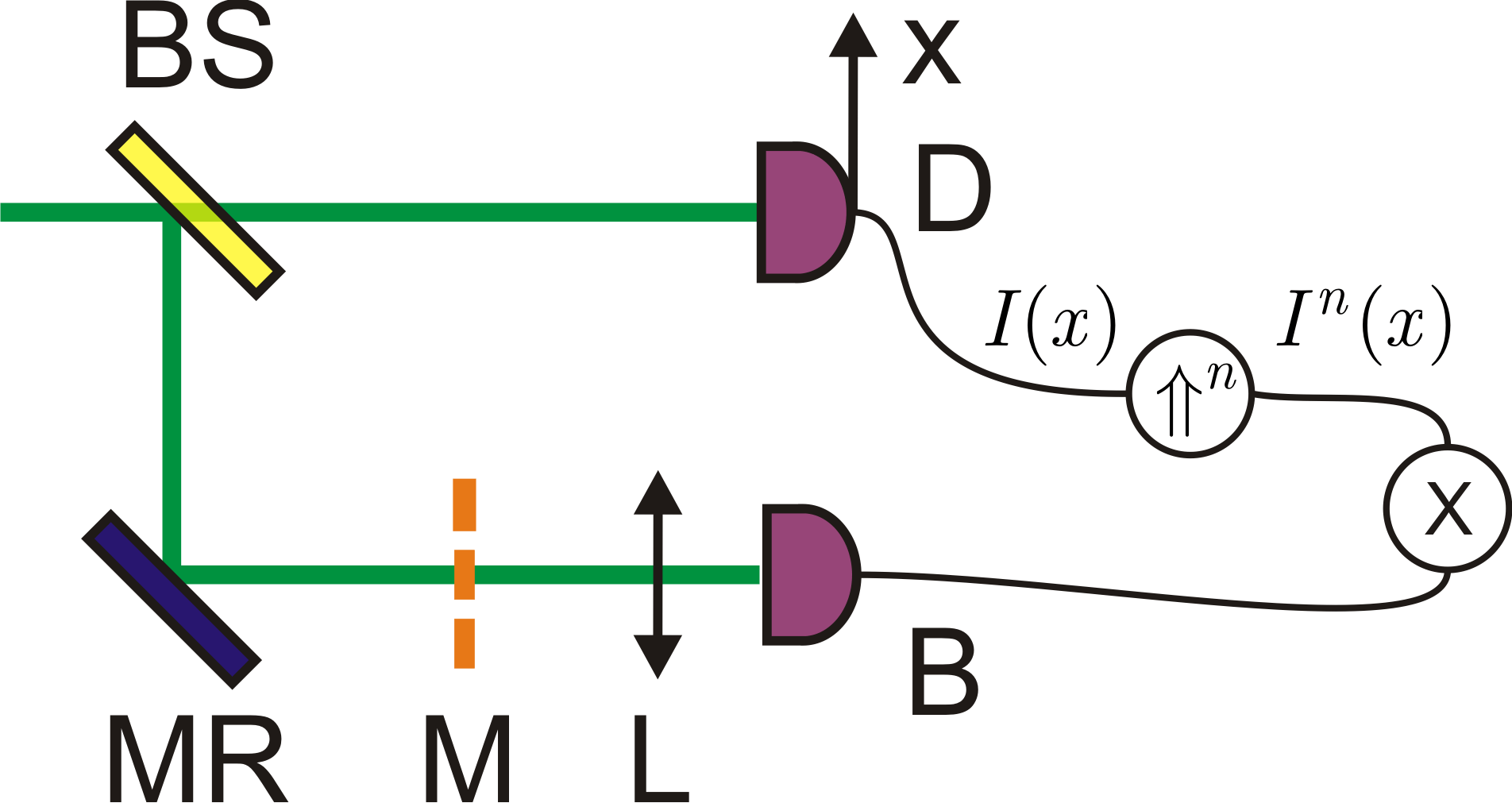} \caption{Schematic of a 2-port HBT setup for
measuring a 2-point n-th order ghost image.}
\end{figure}

Since for thermal light $G^{(i)}$ scales as the i-th order of
intensity, in the high-intensity limit the term with $i=n$ in
(\ref{CF_eq}) dominates, and CFs (\ref{CF_theor}) and
(\ref{CF_meas}) coincide. In this case, our measurement procedure
gives the same as the n-port HBT setup.

\section{Experiment}
Experimental setup is presented in Fig. 5. As the radiation source,
we use a frequency doubled Q-switched Nd:YAG laser with the
wavelength 532 nm, pulse duration 5 ns, and the repetition rate of
47 Hz. The laser beam with diameter set by aperture A$_1$ is
projected onto a rotating ground-glass disk GGD to form
pseudothermal light. The central part of the speckle field is
selected by aperture A$_2$. Polarizing filter P is introduced to
select a linear polarization, thus eliminating any polarization
effects introduced by the mirror. Reflecting from mirrors M1 and M2
the beam is divided into two by a nonpolarizing beam-splitter BS.
Transmitted (signal) and reflected (reference) beams are reflected
by mirrors M3 and M4 respectively. 30-degrees prism PR makes both
beams parallel and sends them to a digital photo camera DPC. Lens L
is used to image the desired area of the speckle field onto the
photo camera's matrix 20.7x13.8 mm (2640x1760 px).

\begin{figure}
\includegraphics[width=0.3
\textwidth]{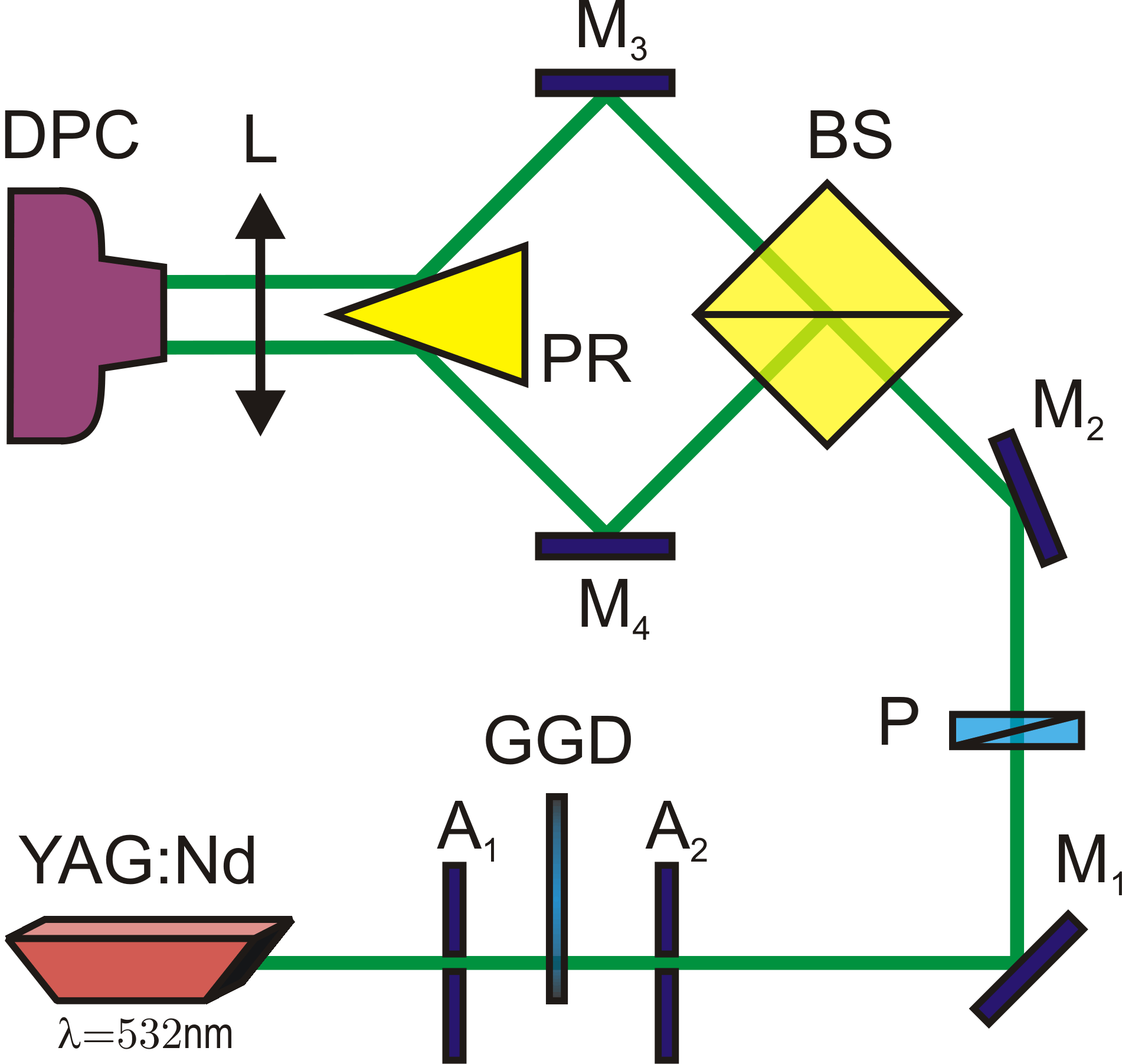} \caption{Experimental setup for measuring n-th
order ghost images. }
\end{figure}

In experiment it is more convenient to use two different parts
(signal and reference) of a single CCD camera to capture both signal
and reference beams instead of using a CCD and a bucket detector as
there is no need to synchronize the latter two. In this case the
signal from the bucket detector is obtained by calculating the total
signal captured by the reference part of the CCD camera. The idea of
using only one CCD camera has been recently implemented in
experiment ~\cite{Basano2}. In our case the ``software'' mask was
calculated using a true ``physical'' realization of the reference
beam instead of using only one beam as both signal and reference.

\begin{figure}
\includegraphics[width=0.3
\textwidth]{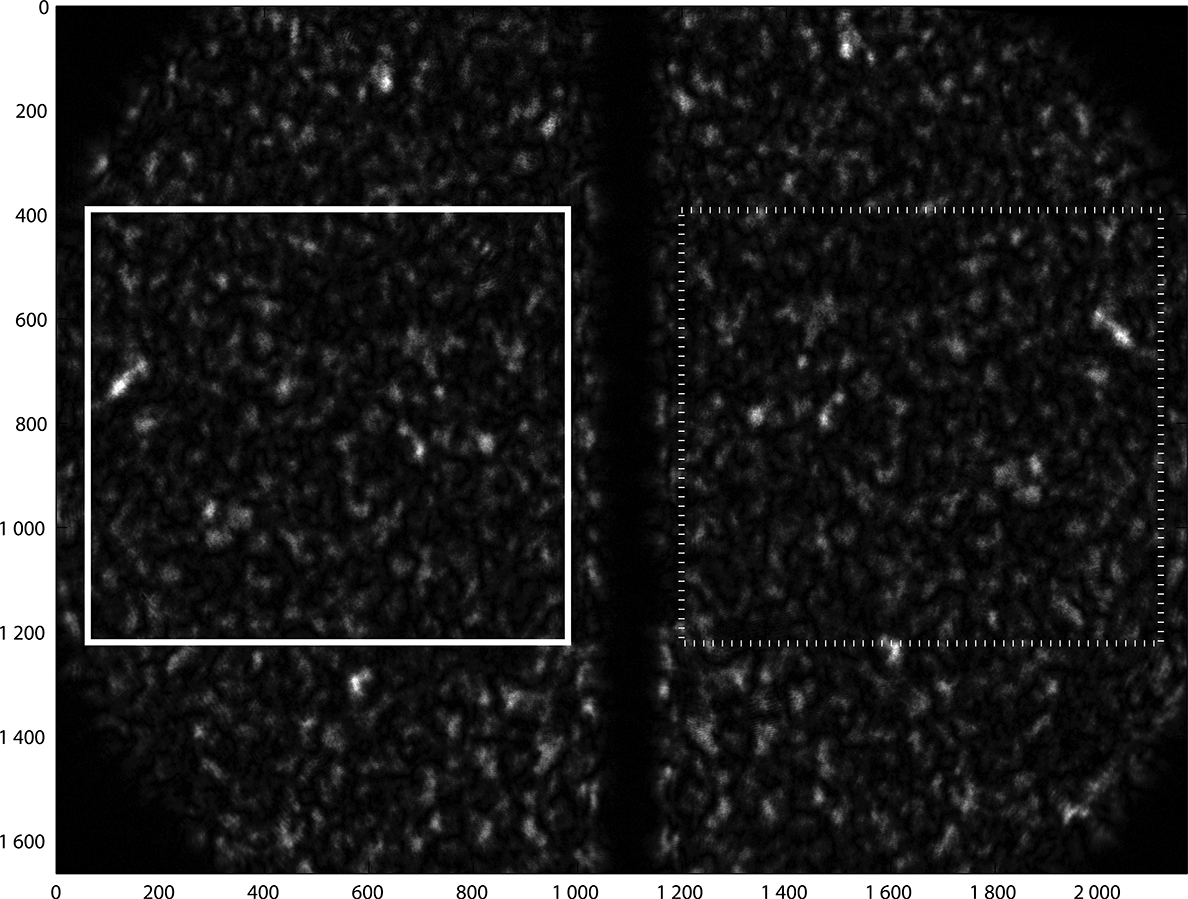} \caption{Single captured frame. The area
marked by the solid and dashed white rectangles represent two
correlated parts of the speckle field that are selected for ghost
image calculation.}
\end{figure}

Exposure time was chosen to be 1/50 seconds, so that only one pulse
was captured at every frame and the frame rate was about 0.2 fps.
Ground glass disk rotation speed was small enough to consider it
stationary for the duration of laser pulse and large enough to allow
considerable (compared to GGD surface inhomogeneity size) linear
displacement between two consequent frames. Thus all the captured
frames were time independent and a total of 5000 frames were
captured. A typical frame is presented in Fig. 6. The average
speckle size was about 30 px (pixels) which correspond to 150$\mu$.
By summing the intensities over a desired area (i.e. mask) we
obtained the ``bucket detector'' signal which was correlated with
every pixel's signal in the signal arm. Two correlated parts of the
signal and reference arms were used for ghost image calculation -
they are depicted by solid and dashed white rectangles in Fig. 6.

\begin{figure}
\includegraphics[width=0.35
\textwidth]{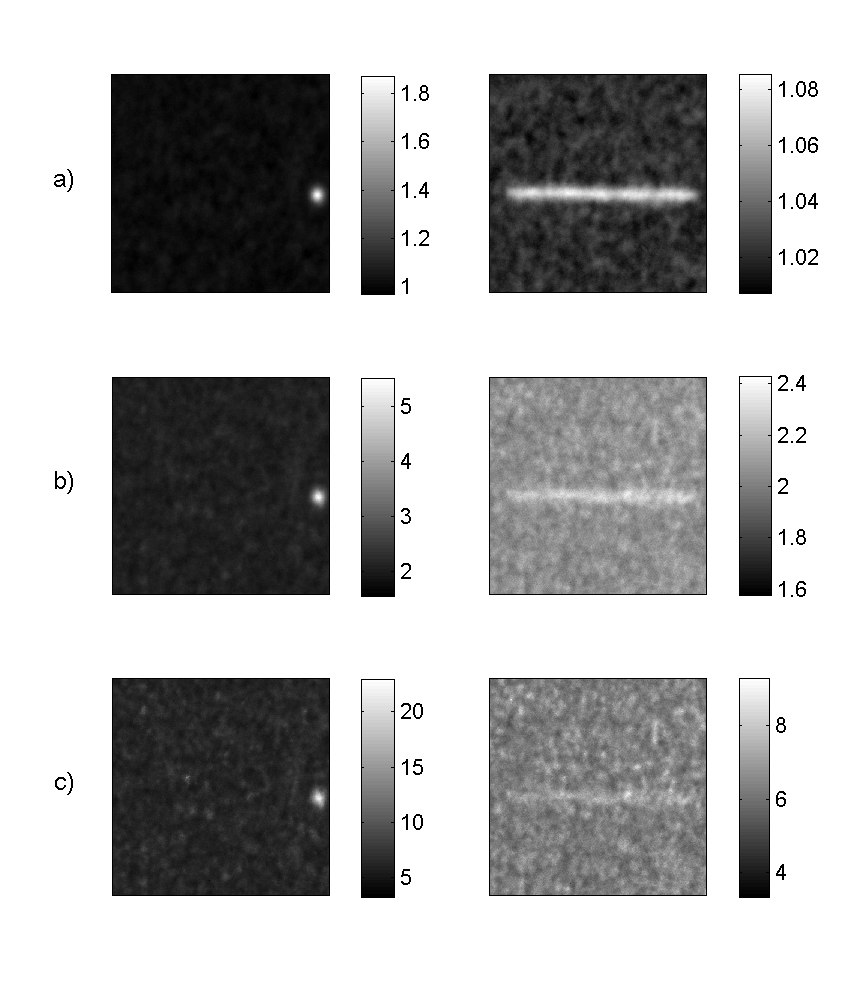} \caption{Ghost images of a short and long
one-dimensional single slits for 2-nd (a), 3-rd (b) and 4-th (c)
order ghost imaging measured using 5000 frames.}
\end{figure}

For clarity we choose a very simple object to image - a single slit
represented by a line of 1 px height and variable width. By changing
the length of the line one can change the number of modes M. An
array of signal values $S(i,j,k,l)$ is calculated as the difference
of $G^{(n)}_{\mathrm{max}}(i,j)$ and $G^{(n)}_{\mathrm{back}}(k,l)$.
Mean value $S$ of $S(i,j,k,l)$ is then calculated as follows
$$S = \left\langle G^{(n)}_{\mathrm{max}}(i,j) \right
\rangle_\mathcal{M} - \left\langle G^{(n)}_{\mathrm{back}}(k,l)
\right\rangle_\mathcal{B},$$ where angle brackets denote averaging
that is done over areas $\mathcal{M}$ and $\mathcal{B}$. The first
area is a line along the the slit with $j=j_{slit}$ (vertical
position of the slit), the latter is a rectangular area of a ghost
image outside of the slit. Noise is calculated as the standard
deviation of $S(i,j_{slit}) = \left\langle S^{(n)}(i,j_{slit},k,l)
\right\rangle_\mathcal{B} $.

As one can see from Fig. 7, SNR (the ``clarity'' of the slit's ghost
image) decreases with the growth of the ghost image order as it is
predicted by the theory. For a bigger object (a longer slit) this
decrease is even more pronounced: the longer slit image has almost
drowned in the noise for the case of fourth-order ghost image. By
changing the size of the slit in steps one can plot the dependence
of visibility and SNR on the size of the mask.

\begin{figure}
\includegraphics[width=0.3
\textwidth]{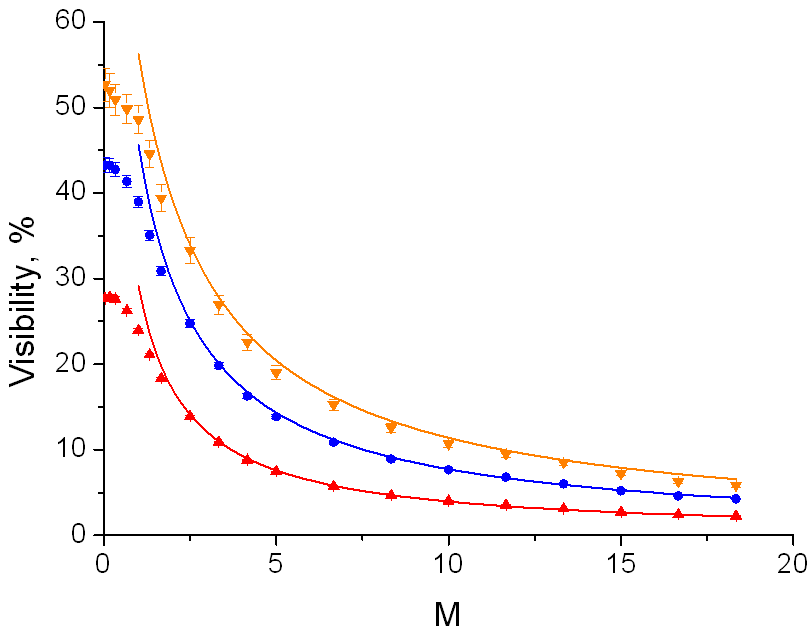} \caption{Visibility dependence on parameter M
(number of modes) for the 2-nd (red), the 3-rd (blue)$\,$and the
4-th (orange)$\,$order ghost imaging of a single slit.}
\end{figure}

From Fig. 8 it is seen that visibility increases for higher-order
ghost images and the values obtained are in a good agreement with
the theoretical fit (solid lines). Due to laser power instability
over time and GGD inhomogeneity, the fit is made by assuming that
the observed values of autocorrelation functions differ from the
theoretical values by constant factors $g^{(n)}=F_n n!$ with their
values close to unity. As far as the ``real'' speckle size might be
different from the one determined as a FWHM, the parameter M is
assumed to be different from a theoretical one by a fixed value $S$.
$F_n$ were independently calculated as mean values of
autocorrelation functions of the corresponding order over 5 million
pixel areas and make up the following values for second-fourth
orders: $F_2=0.988$, $F_3=0.984$ and $F_4=0.997$. The best fit
corresponded to $S=1.19$, which means that the number of modes is
determined by speckle width (FWHM) multiplied by S.

\begin{figure}
\includegraphics[width=0.3
\textwidth]{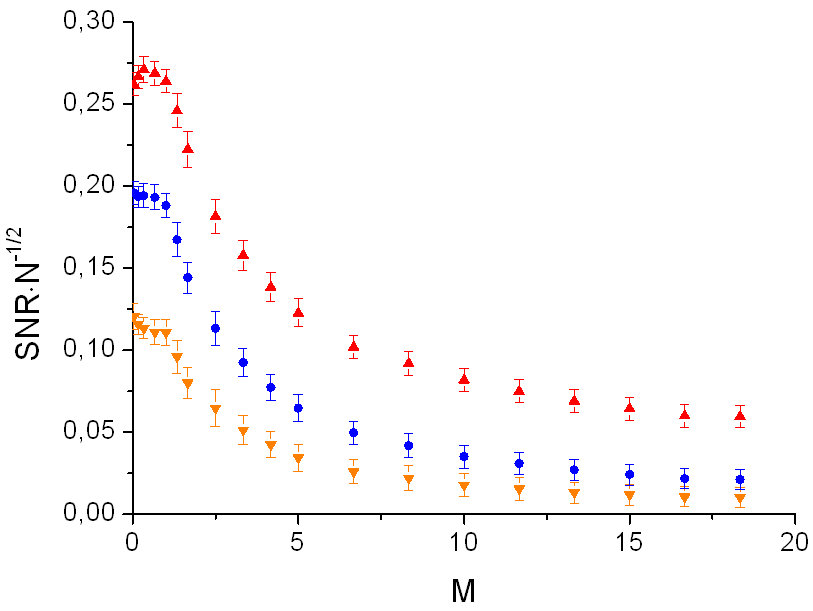} \caption{Normalized SNR dependence on the
parameter M (number of modes) for the 2-nd (red), the 3-rd (blue)and
the 4-th (orange)order ghost imaging of a single slit.}
\end{figure}

In experiment one can only measure sample mean and sample standard
deviation, so, in fact, all SNR values are multiplied by a factor of
$\sqrt{N}$, where N is the total number of images used to calculate
the ghost image. Fig. 9 shows the dependence of the normalized SNR
on the number of modes M. One can see that although the visibility
does increase with the order, the SNR drops. In order to have the
same SNR together with higher visibility values one should increase
the number of images $N$ used.

\section{Conclusion}
We have experimentally performed the 2-nd, the 3-rd and the 4-th
order 2-point ghost imaging in pseudothermal light using a 2-port
HBT setup and theoretically proved that such measurements are
equivalent to those made with 2-, 3- and 4-port setups respectively.
We showed that for higher order ghost images defined as in
expression (\ref{G}) there is indeed an increase in imaging
visibility and decrease in SNR at the same time. It is found that
the second order ghost imaging is optimal for thermal light in the
sense of the highest SNR values compared to those in higher orders
for a fixed sample size (number of images). Higher visibility values
might be obtained using higher order correlations but at the expense
of SNR decrease which, in turn, can be compensated by increasing the
sampling size.

In the single-mode case thermal light has slightly lower SNR values
than SPDC (the difference is less than 6\%). In multi-mode case, the
difference is larger: ghost imaging with SPDC has a higher SNR
maximum compared to thermal light. This result is in agreement with
theoretical consideration in ~\cite{Erkmen2}. But at high
intensities SNR values are the same for both cases at any values of
$M$. From this point of view, if one is interested in best SNR
values, thermal light and SPDC give identical result, whereas if one
is interested in having the best visibility at a fixed number of
images used SPDC has an advantage.

Further research is needed to investigate other cases of ghost
imaging experimentally. For instance, recent results ~\cite{Boyd,
LA} suggest that higher SNR can be obtained by exploiting n-th order
ghost imaging with symmetrical orders of reference and signal
intensities.

This work was supported in part by the RFBR grants \# 08-02-00741,
\# 08-02-00555 and the Program of Leading Scientific Schools
Support, \# NSh-796.2008.2. I.N.A. acknowledges the support of the
'Dynasty' Foundation.


\begin{references}

\bibitem{Shih+Pittman} T.~B.~Pittman, Y.~H.~Shih, D.~V.~Strekalov, and
A.~V.~Sergienko, Phys. Rev. A {\bf 52}, R3429-R3432 (1995).

\bibitem{Meyers} R.~Meyers, K.~Deacon and Y.~Shih, Journal of Modern Optics
{\bf 54}, Nos. 16–17, 2381–2392 (2007).

\bibitem{Han} M.~Zhang, Q.~Wei, X.~Shen, Y.~Liu, H.~Liu, J.~Cheng, and Sh.~Han, Phys. Rev. A {\bf 75}, 021803(R) (2007).

\bibitem{Erkmen1} B.~I.~Erkmen  and J.~H.~Shapiro,  Phys. Rev. A {\bf 78}, 023835 (2008).

\bibitem{Boyd} R.~S.~Bennink,  S.~J.~Bentley, R.~W.~Boyd, Phys. Rev. Lett. {\bf 89}, 113601 (2002).

\bibitem{Lugiato} A.~Gatti, E.~Brambilla, M.~Bache, L.~A.~Lugiato,  Phys. Rev.
Lett. {\bf 93}, 093620 (2004); Phys. Rev. A {\bf 70}, 013802 (2004); F.~Ferri, D.~Magatti, A.~Gatti, M.~Bache, E.~Brambilla, and L.~A.~Lugiato, Phys. Rev. Lett.  94, 183602 (2005).

\bibitem{Shih} G.~Scarcelli, A.~Valencia, and Y.~Shih, Phys. Rev. A 70, 051802(R) (2004);
M.~D'Angelo, A.~Valencia, M.~H.~Rubin, and Y.~Shih, Phys. Rev.A 72, 013810 (2005).

\bibitem{Zhang}D.~Zhang, Y.~H.~Zhai, L.~A.~Wu, X.~H.~Chen, Opt. Lett. {\bf 30}, 2354 (2005).

\bibitem{LA2009}X.~H.~Chen, Q.~Liu, K.~H.~Luo, and L.~A.~Wu,  Opt. Lett. {\bf 34}, 695 (2009).

\bibitem{Lugiatophase}M.~Bache, D.~Magatti, F.~Ferri, A.~Gatti, E.~Brambilla, L.~A.~Lugiato,
Phys. Rev. A {\bf 73}, 053802 (2006).

\bibitem{Basano}L.~Basano, P.~Ottonello, Applied Optics {\bf 46}, 6291 (2007).

\bibitem{ghostJMO} I.~N.~Agafonov, M.~V.~Chekhova, T.~S.~Iskhakov, L.-A.~Wu, Journal of Modern Optics (2009).

\bibitem{LA}X.-H.~Chen, I.~N.~Agafonov, K.-H.~Luo, Q.~Liu, R.~Xian, M.~V.~Chekhova, and L.-A.~Wu, arXiv:0902.3713v1
[quant-ph] (2009).

\bibitem{Erkmen2}B.~I.~Erkmen  and J.~H.~Shapiro, Phys. Rev. A {\bf 79}, 023833
(2009).

\bibitem{Basano2}L.~Basano, P.~Ottonello, OPTICS EXPRESS {\bf 15}, No. 19, 1238617 (2007).

\bibitem{Wick}N.~N.~Bogoliubov and D.~V.~Shirkov, \emph{Quantum Fields}, Benjamin-Cummings Pub.
Co, 1982.

\bibitem{Boyd} K.~W.~C.~Chan, M.~N.~O'Sullivan and R.~W.~Boyd, Opt. Lett. 34, 3343-3345
(2009).
\end{references}
\end{document}